\documentclass[a4paper,11pt]{article}
\usepackage{jinstpub} 
\pdfoutput=1 

\title{Measurement of radon emanation and impurity adsorption from argon gas using ultralow radioactive zeolite}

\author[a,1]{Hiroshi~Ogawa, \note{Corresponding author.}}
\affiliation[a]{CST Nihon University, Surugadai, Kanda, Chiyoda-ku, Tokyo, 180-0011, Japan}
\author[b]{Kenta~Iyoki}
\affiliation[b]{Department of Chemical System Engineering, The University of Tokyo, 7-3-1 Hongo, Bunkyo-ku, Tokyo, 113-8656, Japan}
\author[c]{Minoru~Matsukura}
\affiliation[c]{Institute of Engineering Innovation, The University of Tokyo, 2-11-16 Yayoi, Bunkyo-ku, Tokyo, 113-8656, Japan}
\author[b,c]{Toru~Wakihara}
\author[d,e]{Ko~Abe}
\affiliation[d]{Kamioka Observatory, Institute for Cosmic Ray Research, the University of Tokyo, Higashi-Mozumi, Kamioka, Hida, Gifu, 506-1205, Japan}
\affiliation[e]{Kavli Institute for Physics and Mathematics of the Universe (WPI), the University of Tokyo, Kashiwa, Chiba, 277-8582, Japan}
\author[f]{Kentaro~Miuchi}
\affiliation[f]{Department of Physics, Kobe University, Kobe Hyogo, 657-8501, Japan}
\author[g]{Saori~Umehara}
\affiliation[g]{Research Center for Nuclear Physics (RCNP), Osaka University, Ibaraki, Osaka 567-0047, Japan}

\emailAdd{ogawa.hiroshi@nihon-u.ac.jp}

\abstract{The amount of radioactive impurities contaminated in the detector gases is required to be kept at a very low level for rare event particle physics such as dark matter and neutrino observation experiments. 
Zeolite is a well-known class of materials and is one of the possible candidates for  removing impurities from gases. At the same time, the amount of radioactive impurities released from the adsorbent material needs to be sufficiently small. 
In this paper, a development of a new ultralow radioactive zeolite as a product of the selection of ultralow radioactive materials is reported. Results on 
the radon emanation and impurity adsorption from argon gas measurements are also described.}


\begin{document}
\maketitle
\flushbottom
\renewcommand\thefootnote{\alph{footnote}}
\section{Introduction}
\label{sec:intro}
Dark matter is an undetected elementary particle in the universe. To observe scintillation light and ionization generated by the interaction with dark matter and gas nuclei, dark matter detectors use liquid noble gas (e.g., Xe, Ar, etc) or fluorine compound gas in underground experimental facilities~\cite{P.Gondolo}. Furthermore, ton-scale noble gas detectors have been developed recently. 

One of requirements for such dark matter search experiments is the gas's extremely low impurity and radioactivity. Impurities, including water, can reduce the efficiency of the ionization and scintillation of gases~\cite{MEG_abs}. Furthermore, the events caused by radioactive impurities in the rare gas may not possibly be distinguished from the events caused by dark matter: hence, it is imperative to eliminate the impurities as much as possible. Particularly, $^{222}$Rn emanates from the detector materials and is uniformly distributed in the gas, and impurities must be actively removed.


Zeolite is a possible adsorbent for removing these impurities. Zeolites are known to have about 250 types of structures and these make various adsorption properties. Zeolite A (A-type) has a Si, Al and O backbone structure in which exchangeable cations are present. The pore size of zeolites changes discontinuously from 5~\AA$\rightarrow$~4~\AA$\rightarrow$~3~\AA~when the exchangeable cation is changed from Ca$^{2+}$~$\rightarrow$~Na$^{+}$~$\rightarrow$~K$^{+}$~\cite{S.Upasen}. The specific pore size allows 'molecular sieving', meaning that if the pore size is close to that of the target molecule (radon), the adsorption capacity increases. Adsorption capacity is temperature dependent, and different gases have been discovered to have different adsorption properties at low temperature. These properties enable zeolites to selectively adsorb impurities, and it is popular in environmental and engineering fields. For example, zeolite 13X has been used in particle physics-related experiment to remove impurities such as water in liquid Xe~\cite{MEG_purification}. Furthermore, zeolite 5A has an adsorption capacity for Rn~\cite{Rob}, and it can reduce water and Rn in gases.

However, in detectors that require a low radioactive environment, such as dark matter search and neutrino experiments, the number of radioactive impurities released from zeolite must be reduced to the limit. For example, a radium ($^{226}$Ra) equivalent of 10 mBq / (1-kg zeolite) or less is required in a dark matter search experiment, which is 1/100 to 1/1000 of the commercially available zeolite.

Zeolite has mainly two advantages over other absorbents. First, zeolite for dark matter search experiments can be produced in the laboratory using existing materials and are not expensive, including equipment. Second, zeolites exhibit high efficiency in trapping other impurities such as H$_{2}$O.

Based on the previous result of this study, the synthesis of low-radioactive zeolite was announced by selecting the raw material for a zeolite 4A~\cite{Ogawa1}. In this study, an ultra-low radioactivity 5A zeolite with a pore size suitable for Rn adsorption was synthesized and the performance of impurity removal is reported.
First, Section~\ref{sec:sel} explains the detailed development of the new zeolite 5A. Next, the absorption of impurity by a new zeolite in gas is reported in Section~\ref{sec:perform}.  Then  Section~\ref{sec:concl} provides the conclusion for this study.

\section{Synthesis and preparation of 4A and 5A zeolites}
\label{sec:sel}
Raw materials with low radioactivity must be used to reduce the radioactivity of zeolite. Sodium hydroxide, aluminum hydroxide, and silica components are used for the synthesis of zeolite 4A. 
Furthermore, Ca compound is used for the ion exchange process in type 5A zeolite synthesis. The material for the zeolite solidification is also prepared.
In the previous study, sodium hydroxide from Wako Chemical Corp., aluminum hydroxide BHP39 from Nippon-Light-Metal-Company, Ltd., and silica source snowtex-ST30 from Nissan Chemical Corp. were used as materials for the synthesis of zeolite 4A~\cite{Ogawa1}.
However, although the measurements of the previous two materials reached the upper limit, the silica source still had finite radioactivity of 19.6$\pm$0.3~mBq/kg for $^{226}$Ra. Therefore, for this zeolite synthesis, Fuso Kagakukougyou PL-7 was selected as the silica source and high-purity calcium carbonate provided by a reagent manufacturer through Osaka University was selected as the Ca component. Furthermore, Nissan Chemical's alumina sol AS-200 was newly selected as the solidification material for the molding.



A high-purity germanium (HPGe) detector at the Institute for Cosmic Ray Research, University of Tokyo was used to screen these materials. Results are shown in Table~1. No significant radiation was reported for all materials used and 90~\% upper limits are shown in the table.  
Furthermore, ultrapure water was provided by Organo Corporation for the synthesis and cleaning. TAMAPURE-AA from Tama Chemical Co., Ltd., was used as nitric acid for dissolving calcium carbonate. The uranium concentrations for ultrapure water and nitric acid were $<$2~ppt and $<$10~ppt in the elemental impurity measurements using inductively-coupled-plasma-mass spectrometry, which was sufficiently small. 
\begin{table}[htbp]
\begin{center}
\caption{Materials selected for this zeolite synthesis and their radioactivity measured with a HPGe detector.}
\begin{tabular}{lccl}
    \hline \hline
    Material&$^{226}$Ra[mBq/kg]&$^{232}$Th[mBq/kg]&Company/Commercial name\\
    \hline  
   NaOH&$<$12.2&$<$8.1&WAKO/NaOH for precise analysis\\
   Al(OH)$_{3}$&$<$9.1&$<$4.3&Nihon Light Metal Company Ltd./BHP39 \\
   Silica source&$<$5.8&$<$4.6&Fuso Kagakukougyou/Colloidal silica PL-7\\
   Ca compound&$<$17.0&$<$6.6&Osaka-U/calcium carbonate\\
   solidification mat.&$<$4.3&$<$4.2&Nissan kagaku/Alumina sol AS-200\\
    \hline \hline
\end{tabular}
\end{center}
\label{tabselRI}
\end{table}

A trial synthesis of zeolite using these materials was conducted in collaboration with Nihon University College of Science and Technology and the University of Tokyo. A clean booth with an ultralow-penetration-air filter was set up in the laboratory of Nihon University College of Science and Technology Funabashi Campus to prevent contamination from the outside, where the synthesis was taken place.

First, the manufacturing process of 4A-type zeolite is described. Aluminate was prepared by mixing sodium hydroxide, aluminum hydroxide and pure water. Then, it was synthesized with dissolved colloidal silica, aluminate, an aqueous solution of sodium hydroxide, and pure water. The synthesized solution was heated in a constant temperature bath for the crystallization, and the product was washed with pure water, and then dried in a constant temperature bath to remove water. 

Second, solidification process of zeolite is described. Alumina sol was mixed with the synthesized zeolite 4A. Pure water was added to reduce the viscosity. The mixture was placed thinly on a stainless steel tray. Then, it was dried to solidify in an electric furnace. The solidified zeolite was crushed and sieved to an appropriate size, which completes the solidified zeolite 4A.

\begin{figure}[htbp]
  \begin{center}
    \includegraphics[keepaspectratio=true,height=50mm]{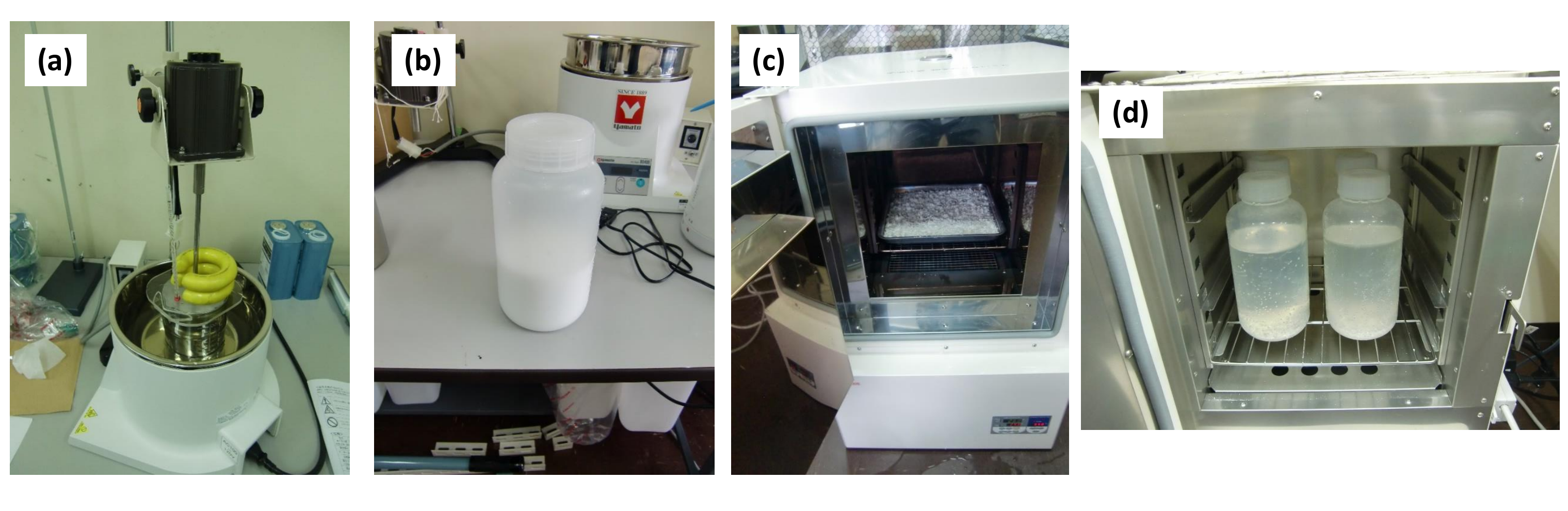}
  \end{center}
  \caption{Development process of a zeolite 5A. The images indicate (a) aluminate making,  (b) silica and aluminate mixing before the crystalization for 4A type, (c) solidification process and (d) ion exchange process.}
  \label{fig:pic-zeo}
\end{figure}

Finally, the ion exchange is described. Calcium carbonate was mixed with dilute nitric acid to synthesize an aqueous calcium nitrate solution. A solidified zeolite 4A was placed inside and maintained in a constant temperature bath to exchange the Na ions with Ca ions. Then, it was washed with pure water and dried to complete the 5A-type zeolite. Figure~\ref{fig:pic-zeo} shows the development process of zeolite 5A; the weight of the zeolite produced in Fig.~\ref{fig:pic-zeo} is approximately 200~g.


The radioactivity of the synthesized zeolite 5A was measured using a HPGe detector. The value of radioactivities is 14.2 $\pm$ 7.0 mBq/kg for $^{226}$Ra and 58.8 $\pm$ 8.0 mBq/kg for $^{232}$Th.  
This result is consistent with the value of radioactivities contained in the materials within the upper limit for $^{226}$Ra. 
However, in the case of $^{232}$Th, there is a possibility of contamination during the fabrication process, unrelated to the materials themselves.
Table~\ref{tababsRI} shows a comparison of radio-iotopes (RI) measurement results for this work and other adsorbents.
The $^{226}$Ra contamination of the synthesized zeolite 5A was found to be 0.3~$\%$ or less of that of a commercial zeolite, which was measured to be 5.5 Bq/kg.

\begin{table}[htbp]
\begin{center}
\caption{The comparison of RI measurement results for this work and other adsorbents. RI contaminations were measured with Ge detectors, except for Carbo-Act, which is the result of neutron activation analysis; the Carbo-Act value is considered to be the result of a decay equilibrium between $^{238}$U and $^{226}$Ra.}
\begin{tabular}{lcc}
    \hline \hline
    Absorbents&$^{226}$Ra[mBq/kg]&$^{232}$Th[mBq/kg]\\
    \hline
    Selecto 32-63$\mu$m, Silica gel acid washed~\cite{SNO}&566.7$\pm$46.8&256.4$\pm$30.9\\
    Carbo-Act Int. activated charcoal~\cite{SNO}&2.5$\pm$0,3&0.5$\pm$0.1\\
    Shirasagi G2x 4/6 activated charcoal~\cite{XMASS}&67$\pm$15&--\\
    Unitika Ltd. activated carbon fiber A-20~\cite{Nakano}&$<$5.5&$<$10.4\\
    \hline
    Commercial Molecular sieves 4A&5500$\pm$200&7000$\pm$100\\
    4A Zeolite solidifying~\cite{Ogawa1} &57.0$\pm$14.0&198.4$\pm$16.5\\
    5A Zeolite (This work)&14.2$\pm$7.0&58.8$\pm$8.0\\
    \hline \hline
\end{tabular}
\end{center}
\label{tababsRI}
\end{table}

\section{Performances of the zeolite}
\label{sec:perform}
A Rn emanation measurement and an impurity adsorption test were performed for the zeolite synthesized in this work. First, the filters for the emanation and impurity measurements are explained. Furthermore, Rn measurement systems are explained. Then, the results of Rn emanation, Rn removal, and water removal from Ar gas by the synthesized zeolite are shown.
\subsection{Zeolite filter} 
\begin{figure}[htbp]
  \begin{center}
    \includegraphics[keepaspectratio=true,height=65mm]{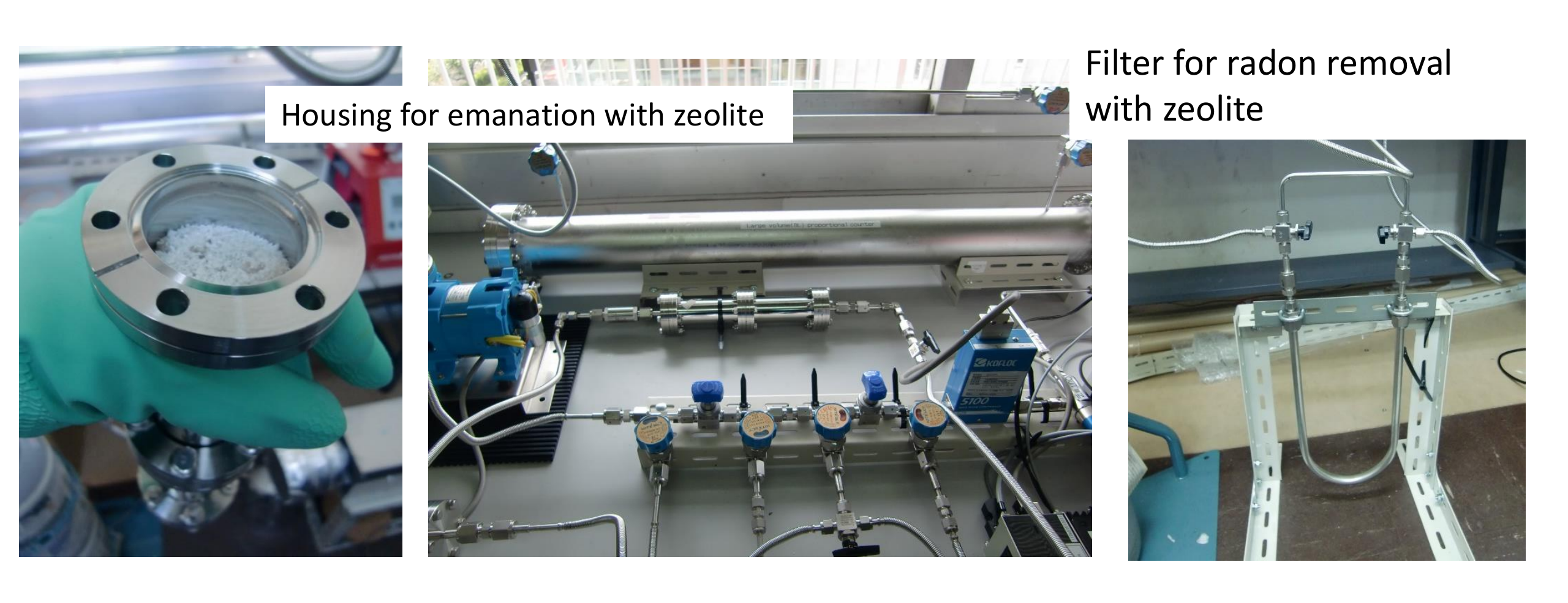}
  \end{center}
  \caption{Photos of the zeolite used as a gas filter. The left photo shows how the zeolite is seen in the housing. The center and the right photos show the ICF nipple and a bent stainless pipe, respectively. See the gas line diagram in Fig.~\ref{fig:setup1} for the respective filter placement.}
  \label{fig:filter}
\end{figure}

The synthesized zeolite is usually packed in a housing when it is used as a gas filter, and their photos are shown in Figure~\ref{fig:filter}. For the Rn emanation test, 172~g of zeolite was placed in an ICF nipple, while, 
for the impurity adsorption test, 39~g of zeolite was placed in a filter housing made of a bent stainless pipe, half-inch in diameter and 25 inches in length by following procedure. Both housings have a metal nanoparticle filter in both ends. The filter was activated before the test. First, the zeolite was heated to 200~$^{\circ}$C in an oven to remove the adsorbed water. Then it was placed in a housing and the housing was heated with a ribbon heater at 200~$^{\circ}$C for one hour,  while inside of the housing was purged with nitrogen gas. Then, the filer was set in a test line and was heated again at 200~$^{\circ}$C for one hour while the test line was being evacuated.

\subsection{Radioactivity measurement system}
\begin{figure}[htbp]
  \begin{center}
	 \includegraphics[keepaspectratio=true,height=80mm]{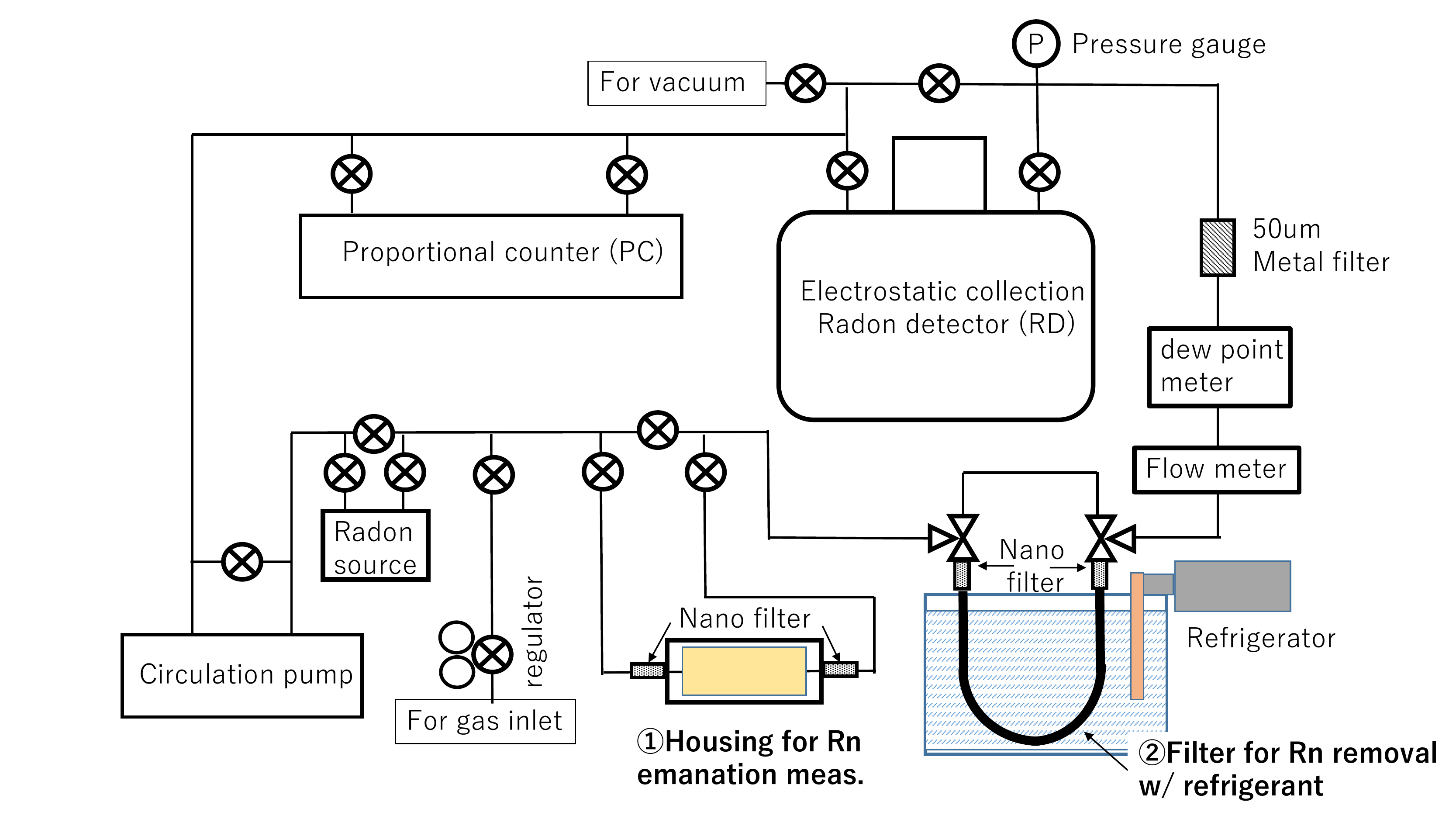}
    \includegraphics[keepaspectratio=true,height=80mm]{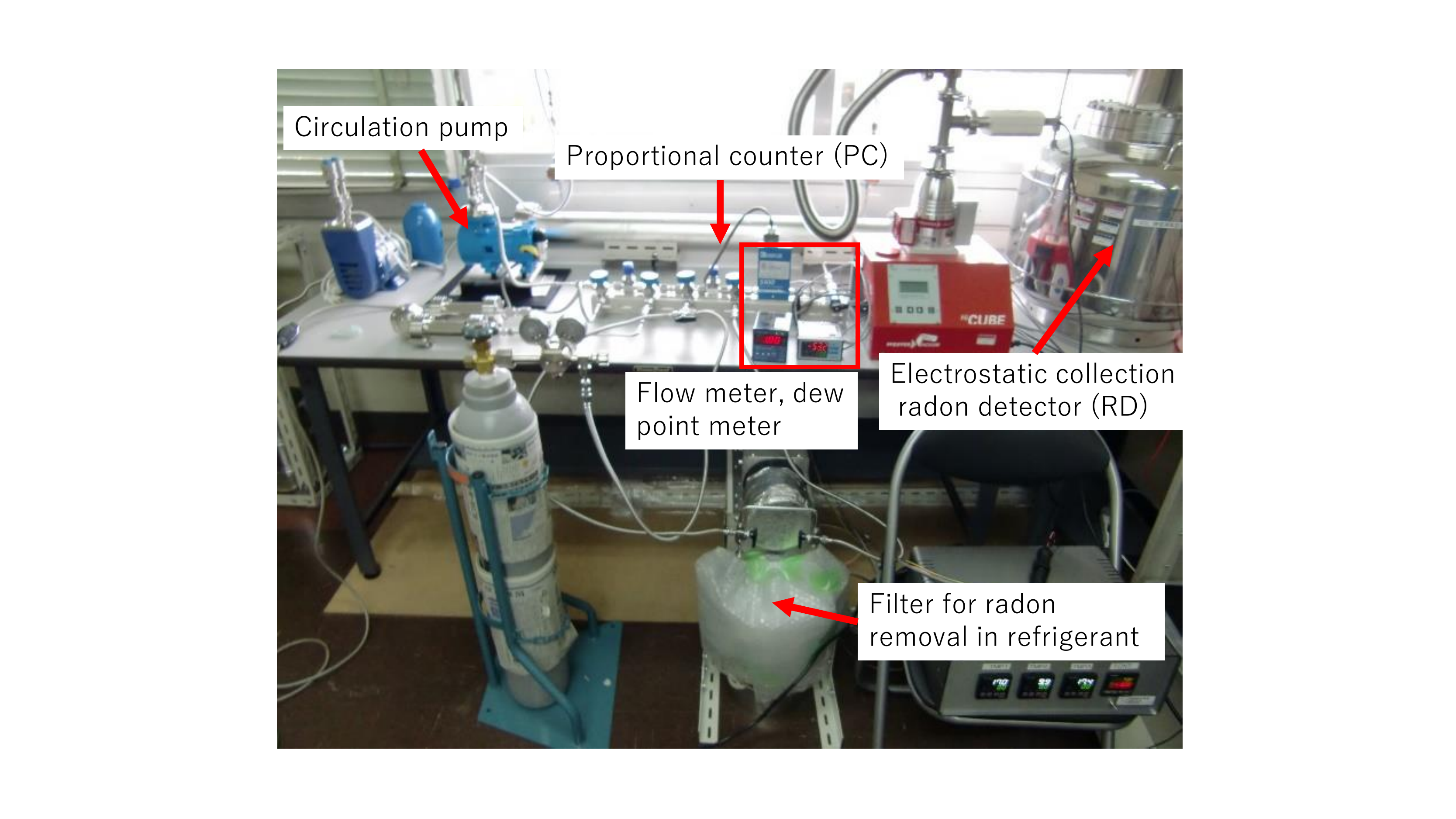}
  \end{center}
  \caption{(Top) A schematic representation of the gas line and the radioactivity measurement system. (Bottom) An overview photo of the radioactivity measurement system.}
  \label{fig:setup1}
\end{figure}

The radioactivity measurement system used in this study is described in detail in~\cite{Ogawa2}. There exist two types of radon detectors.
One is the electrostatic collection radon detector (RD) which alpha-rays from $^{214}$Po, the daughter nucleus of $^{222}$Rn, are measured.
The other is a proportional counter (PC) to observe alpha-rays that decay and emit from $^{222}$Rn directly. 
A circulation pump was used to circulate the gas. The gas circulation speed was set at 1 L/min. A flow meter, a pressure gauge, and a dewpoint meter monitoring the moisture in gas are equipped in the system. Each of the Rn data is recorded with a multichannel analyzer. A refrigerator and a refrigerant (3M$^{\rm TM}$ Novec$^{\rm TM}$ 7200) can cool the impurity adsorption test housing. Figure~\ref{fig:setup1} provides a schematic representation of gas line and the radioactivity measurement system, as well as an overview photo of the setup employed for this study.
\subsection{Radon emanation measurement}
\begin{figure}[htbp]
  \begin{center}
    \includegraphics[keepaspectratio=true,height=60mm]{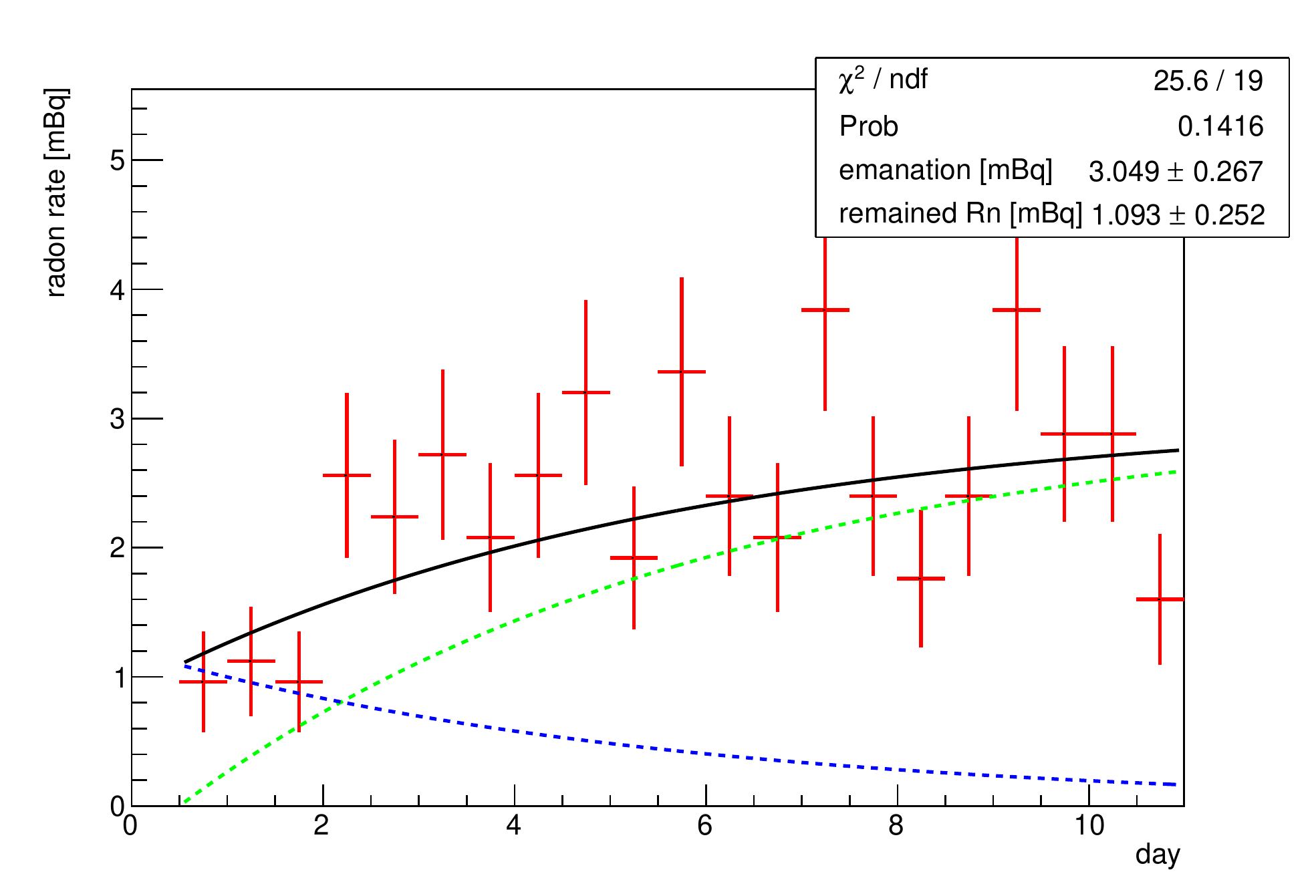}
  \end{center}
  \caption{Time profile of the measured radon rate (red) for the Rn emanation measurement. Best-fit Rn emanation (green), remained Rn (blue) and total (black) rates are also shown.}
  \label{fig:emanation}
\end{figure}

The Rn emanation was measured for the solidified zeolite 5A. Ar gas was circulated from the zeolite filter and RD at room temperature. The Rn rate is estimated from the count rate in an energy region of $^{214}$Po in the alpha-ray spectrum taking into account the detection efficiency 
which depends on the dewpoint of gas. Measured time dependence of the Rn decay rate for the 5A zeolite is shown with red histograms in Figure~\ref{fig:emanation}. Statistic errors are shown with error bars. The time profile of the Rn rate was fitted as follows;
\begin{equation}
R=A\cdot(1-e^{-t/\tau})+B\cdot e^{-t/\tau}, 
\label{eq:fitting}
\end{equation}  
where $R$ is the observed Rn rate. $A$ and $B$ are the Rn emanation rate from the material and the remaining Rn rate in the detector at the beginning of the measurement, respectively. $\tau ( = 3.82/ln2 = 5.52$ days) is the lifetime of $^{222}$Rn, and $t$ is the time from the start of the measurement. The Rn emanation rate ($A$) was estimated to be 3.0$\pm$0.3~mBq. This value corresponds to 17.4$\pm$1.7 mBq/kg and is consistent with $^{226}$Ra concentration predicted by the HPGe detector measurement.

\subsection{Rn removal measurement}
\begin{figure}[htbp]
  \begin{center}
    \includegraphics[keepaspectratio=true,height=60mm]{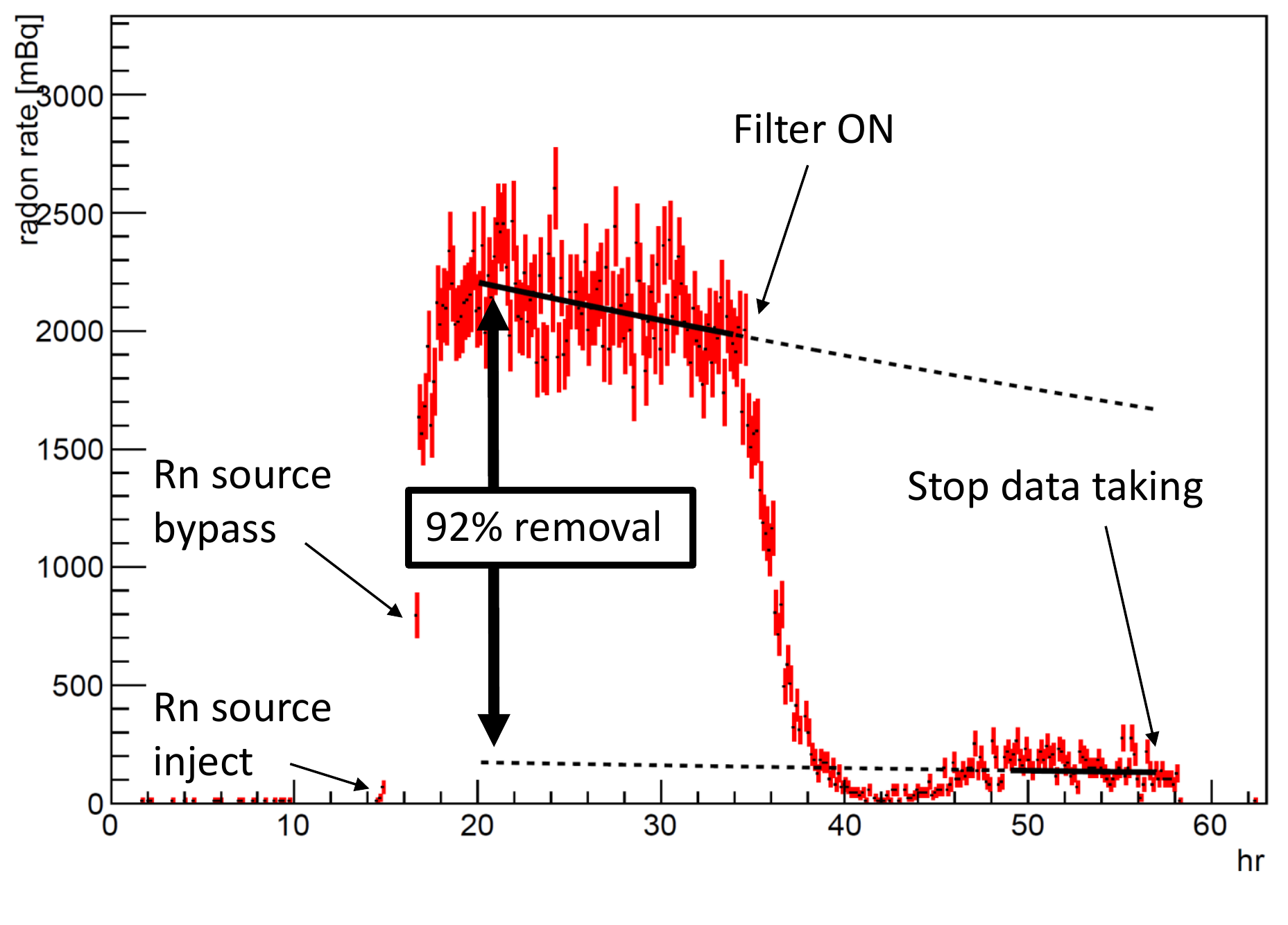}
  \end{center}
  \caption{Time profile of the radon rate during the Rn removal test measured with the RD.}
  \label{fig:rnremoval}
\end{figure}

Rn removal was measured using the RD and the PC. The time profile of the Rn rate during the Rn removal test is shown in Figure~\ref{fig:rnremoval}. Rn was injected from a Rn source (Ra) by passing Ar gas. After bypassing the Rn source, Ar gas was circulated for 18 h at 1L/min to ensure a stable Rn rate, estimated at the decay of $^{214}$Po alpha. Then the Ar gas was redirected to go through the filter (Filter ON) cooled at -60~$^{\circ}$C.
The Rn rate was rapidly reduced to approximately zero in the first phase of filter ON. Thereafter, the radon rate increased slightly due to diffusion effects.
The time profile of the Rn rate was fitted before and after filter ON as follows:
\begin{equation}
R=C\cdot e^{-(t-t_{0})/\tau}
\label{eq:fitting2}
\end{equation}
where $t_{0}$ denotes the time the reference's starting point is located. The reduction rate before and after Rn removal was estimated using the values $C$ of before and after the filter ON as 92 $\%$.
 
\begin{figure}[htbp]
  \begin{center}
	\includegraphics[keepaspectratio=true,height=60mm]{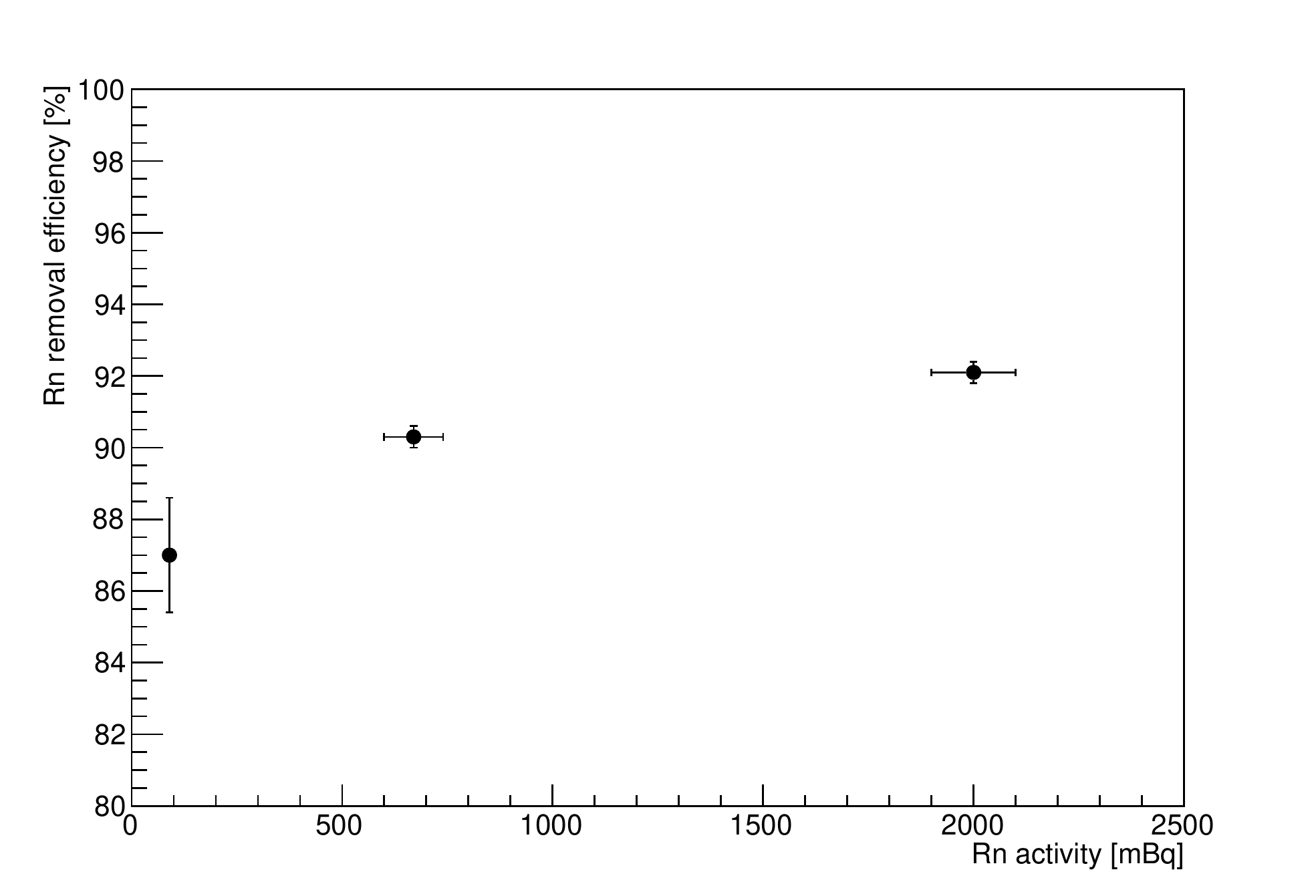}
 \end{center}
  \caption{Results of the radon removal measurements at three radon activities immidediately before applying the filter ON.}
  \label{fig:rndensdep}
\end{figure}

The radon removal measurements were conducted using the specified flow rate of 1~L/min, filter temperature of -60~$^{\circ}$C, 39~g of zeolite and argon gas as a reference.
The dependence of radon removal efficiency on radon activity was investigated, as shown in Fig.~\ref{fig:rndensdep}. The results show that the removal efficiency is close to constant between 500 and 2500~mBq and it is slightly lower at 100~mBq. In the subsequent measurements, the radon activity in the gas varied depending on the accumulation of radon in the Ra ceramic ball. As the measurements were made at activities of 500-2500~mBq, This effect introduced a systematic error of 1.8~$\%$ for the rate of radon removal efficiency.
Table~\ref{tab:removaltest} summarizes the results of radon removal efficiency with various flow rates, filter temperatures, zeolite contents and gas species (from argon to air). The findings suggest that the adsorption efficiency decreases slightly with a flow rate increase, but the difference is not significant within the test range. The removal efficiency shows a significant dependence on the filter temperature, and it is evident that the efficiency decreases with a lower mass of zeolite. Additionally, the removal efficiency for radon in air is lower than that in argon. However, by lowering the filter temperature to -96~$^{\circ}$C, the removal efficiency for air can be improved. 

\begin{figure}[htbp]
  \begin{center}
	\includegraphics[keepaspectratio=true,height=80mm]{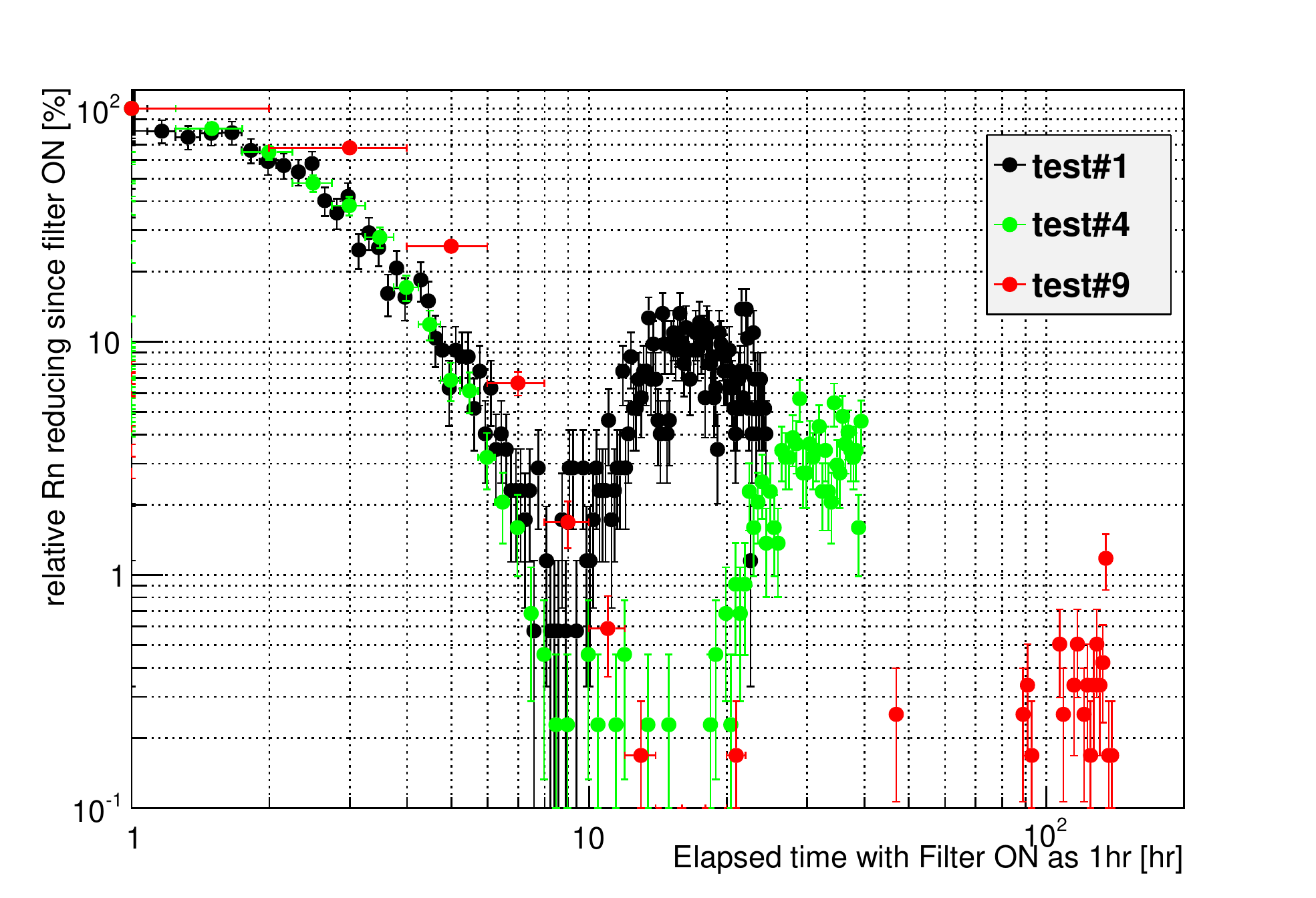}
 \end{center}
  \caption{Time profile of the radon reduction factor in the three adsorption tests in Table~\ref{tab:removaltest} (test~$\#$1, test~$\#$4 and test~$\#$9). The time and Rn reducing are scaled by the time and value corresponding to the filter ON.}
  \label{fig:rnlimit}
\end{figure}

In order to evaluate the limits of radon adsorption capacity, several additional tests were conducted. For Ar gas, the amount of zeolite was doubled by connecting zeolite filters of the same type in series. The temperature was cooled to -96~$^{\circ}$C and the flow rate was reduced to 0.5~L/min. This test, referred to as test~$\#$9, was compared to a reference adsorption test (test~$\#$1) and the most efficient case (test~$\#$4) at a temperature of -96~$^{\circ}$C. Figure~\ref{fig:rnlimit} presents the time profiles of radon reduction factor in the three adsorption tests. In test~$\#$9, when Ar gas was introduced into the filter, almost no radon was observed for approximately 70~h, following by a breakthrough. The time until breakthrough was 2~h in test~$\#$1 and 10~h in test~$\#$4. This indicates that not only did the removal efficiency increase, but the breakthrough time also extended, demonstrating an improvement in the adsorption capacity.

\begin{figure}[htbp]
  \begin{center}
    \includegraphics[keepaspectratio=true,height=80mm]{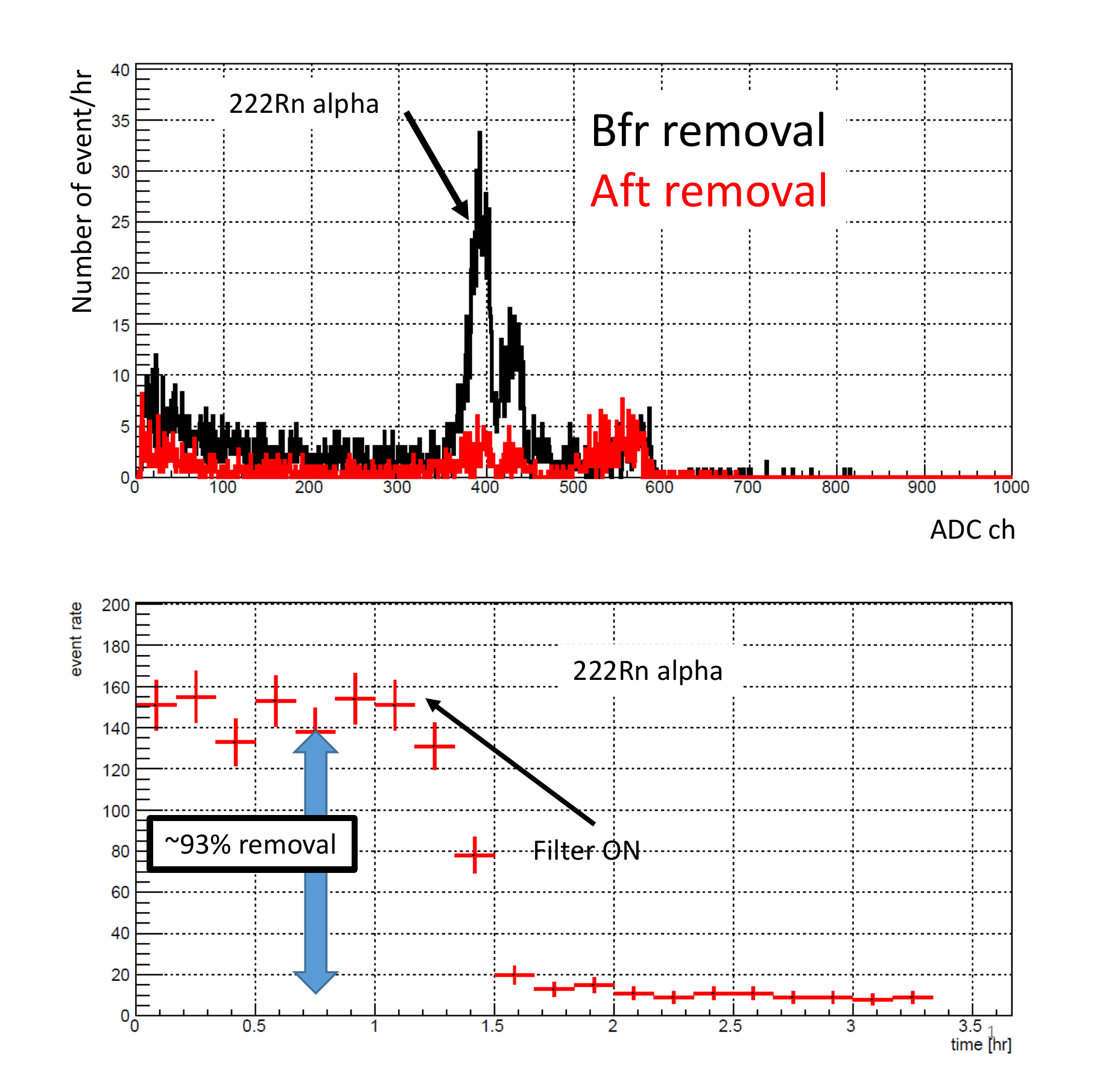}
  \end{center}
  \caption{Observed energy spectra before and after the filter ON and the time profile of the radon rate during the Rn removal test measured with the PC.}
  \label{fig:rnremoval_prop}
\end{figure}

The absorption of $^{222}$Rn by the filter was also confirmed with a PC. 
Observed spectra before and after filter ON and the time profile of the Rn removal test using the PC are shown in Figure~\ref{fig:rnremoval_prop}. The reduction of the peak of $~^{222}$Rn alpha-rays after filter ON is observed. The reduction rate of $^{222}$Rn is 93~$\%$, which agrees with the result of the RD. 

\begin{table}[h]
\begin{center}
\caption{Results of radon removal efficiency with various flow rates, filter temperatures, zeolite contents and gas species (from argon to air). }
\begin{tabular}{lccccc}
    \hline \hline
    Test ID&Flow rate&Temperature&Mass&Gas&Removal\\
	 &[L/min]&[$^{\circ}$C]&[g]&&efficiency [$\%$]\\
    \hline
   $\#$1&1&-60&39&Argon&92.1$\pm$1.8\\
    $\#$2&0.5&-60&39&Argon&91.1$\pm$1.8\\
    $\#$3&2&-60&39&Argon&88.2$\pm$1.8\\
    $\#$4&1&-96&39&Argon&96.6$\pm$1.8\\
    $\#$5&1&0&39&Argon&22.8$\pm$2.3\\
    $\#$6&1&-60&7&Argon&72.8$\pm$1.9\\
    $\#$7&0.7&-60&39&Air&65.6$\pm$1.9\\
    $\#$8&0.7&-96&39&Air&89.2$\pm$1.8\\
    $\#$9&0.5&-96&78&Argon&99.4$\pm$1.8\\
    \hline \hline
\end{tabular}
\end{center}
\label{tab:removaltest}
\end{table}
\subsection {Water removal}
\begin{figure}[htbp]
  \begin{center}
    \includegraphics[keepaspectratio=true,height=60mm]{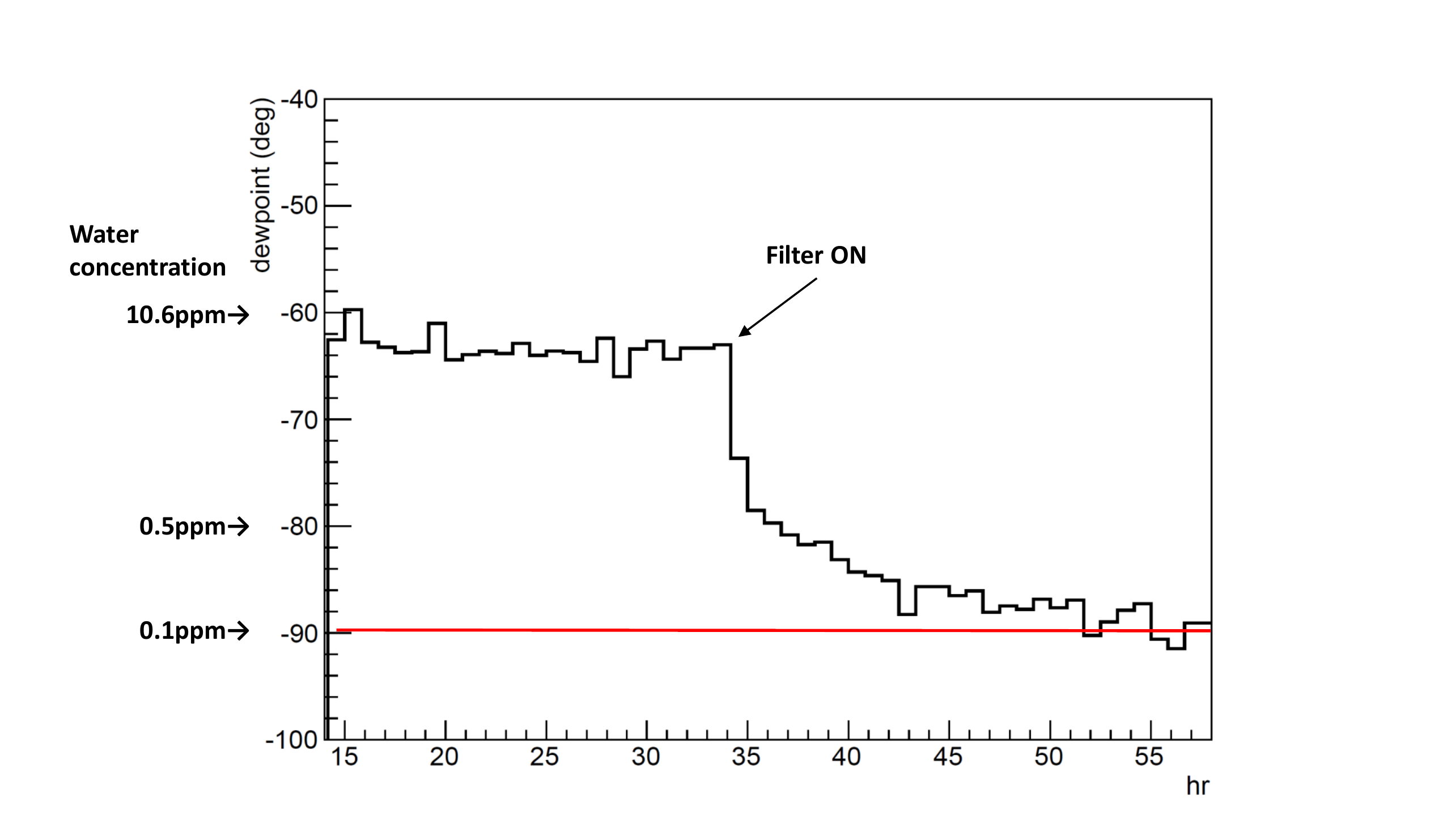}
  \end{center}
  \caption{Time profile of the water concentratin in the system measured with a dewpoint meter. The values converted from the dew point to the water concentration are shown.}
  \label{fig:waterremoval}
\end{figure}

Water removal during filter passing was then performed. Figure~\ref{fig:waterremoval} shows the time profile of the dew point or the water concentration in the system using a dewpoint meter. The dew point meter is a TEKHNE Corporation TK-100. This dew point meter has a measuring range of 20~$^{\circ}$C to -100~$^{\circ}$C and is calibrated from 10~$^{\circ}$C to -80~$^{\circ}$C. Before the filter ON, Ar dewpoint stayed at -65~$^{\circ}$C with a water concentration of 10~ppm. After filter ON, the dewpoint was reduced to -80~$^{\circ}$C in a few hours and reached -90~$^{\circ}$C, corresopnding to the concentration of about 0.1~ppm. 
\section{Conclusion}
\label{sec:concl}
A zeolite that can be used to remove impurities 
in an extremely low background environment 
for rare event particle physics experiments was developed in this work.
The $^{226}$Ra radioactive impurities in this zeolite were measured to 14.2$\pm$7.0~mBq/kg, which corresponds to 0.3~$\%$ of the commercially available zeolite. Measured radon emanation was 17.4$\pm$1.7 mBq/kg.
We performed a radioactive impurity adsorption test on the synthesized zeolite. The results indicate that reducing the temperature significantly improves the efficiency of radon removal. Furthermore, the adsorption efficiency is influenced by the filter temperature and the amount of zeolite used, while the impact of the flow rate is not significant in our test range. These findings will be valuable for the design of radon removal filter systems in dark matter search experiments. The removal efficiency of air was also found to be poor according to argon, but this too showed an improvement in adsorption efficiency by lowering the temperature.
Also water removal down to around 0.1 ppm were achieved from argon gas.
\section*{Acknowledgments}
We acknowledge the support of XMASS collaboration.  
We gratefully acknowledge ORGANO Co.,Ltd and the cooperation of Kamioka Mining and Smelting Company. 
This work was supported by the Japanese Ministry of Education, Grant-in-Aid for Scientific Research on Innovative Areas Region No. 2603 (19H05805 and 19H05806), JSPS KAKENHI Grant No. 19K03893, 16H02189, 18H03699, 22H00127, 23K03435, Foundation for Precision Measurement Technology (2019), Research grants Foundation for CST Nihon University (2020), Academic Award Scholarship for CST Nihon University (2022), Kyoudou-riyou for Institute for Cosmic Ray Research  University of Tokyo.

\end{document}